\documentclass[epj,final]{svjour}

\usepackage{graphicx}
\usepackage{dsfont}
\usepackage{amsfonts}
\usepackage{amsmath}
\usepackage{hyperref}

\graphicspath{{../images/}}

\newcommand{\expect}[1]{\langle {#1} \rangle}
\newcommand{\Tr}[1]{\operatorname{Tr}\big\{ {#1} \big\}}
\newcommand{\komm}[2]{\big[#1,#2\big]}

\allowdisplaybreaks

\begin{document}

\title{Generalized mean-field approach to simulate the dynamics of large open spin ensembles with long range interactions}
\titlerunning{Generalized mean-field approach}

\author{S. Kr\"amer\inst{1} and H. Ritsch\inst{1}}
\institute{Institute for Theoretical Physics, Universit\"at Innsbruck, Technikerstrasse 21, 6020 Innsbruck, Austria}

\abstract{
We numerically study the collective coherent and dissipative dynamics in spin lattices with long range interactions in one, two and three dimensions. For generic geometric configurations with a small spin number, which are fully solvable numerically, we show that a dynamical mean-field approach based upon a spatial factorization of the density operator often gives a surprisingly accurate representation of the collective dynamics. Including all pair correlations at any distance in the spirit of a second order cumulant expansion improves the numerical accuracy by at least one order of magnitude. We then apply this truncated expansion method to simulate large numbers of spins from about ten in the case of the full quantum model, a few thousand, if all pair correlations are included , up to several ten-thousands in the mean-field approximation. We find collective modifications of the spin dynamics in surprisingly large system sizes. In 3D, the mutual interaction strength does not converge to a desired accuracy within the maximum system sizes we can currently implement. Extensive numerical tests help in identifying interaction strengths and geometric configurations where our approximations perform well and allow us to state fairly simple error estimates. By simulating systems of increasing size we show that in one and two dimensions we can include as many spins as needed to capture the properties of infinite size systems with high accuracy. As a practical application our approach is well suited to provide error estimates for atomic clock setups or super radiant lasers using magic wavelength optical lattices.
}

\maketitle

\section{Introduction}
Ensembles of interacting spins in various geometries have been at the heart of quantum statistical physics since the first models on magnetism were proposed~\cite{heisenberg1928theorie}. As the spin-spin interaction is nonlinear and the corresponding Hilbert space grows exponentially with the number of spins, exact analytic as well as full numeric solutions are only possible for very special cases and geometries~\cite{thacker1981exact,fisher1964magnetism,haldane1983nonlinear} or a small number of spins. The complexity increases even further for an open system including collective decay. As a very successful approximate numerical approach based on the factorization of single site expectation values, a dynamical mean-field method was developed for efficient treatment of larger systems~\cite{van1990exact}. On the one hand it allowed for analytical results in the large dimensions limit~\cite{pearce1978} while, on the other hand, it soon proved very useful for a numerical treatment in low dimensions. Subsequently, the general idea of the method was successfully applied to a wide range of solid state physics models in the very low temperature quantum domain~\cite{georges1996dynamical}. Recently, this approach has proven useful in the description of ultra-cold particle dynamics in optical lattices~\cite{byczuk2008correlated,altman2003phase}.

The present work is motivated by another, more recent implementation of spin lattices based on ultra-cold atoms or molecules trapped in an optical lattice, which nowadays can be prepared almost routinely in the laboratory with well defined filling factors and close to zero temperature~\cite{lewenstein2007ultracold,micheli2006toolbox}. When excited on an optical or infrared transition, the trapped particles will interact via dipole-dipole energy exchange forming collective excitations~\cite{ficek1986cooperative,zoubi2013excitons}. In addition, optical transitions intrinsically exhibit dissipation via spontaneous decay, which in such a lattice becomes a collective effect leading to super- or sub-radiance~\cite{zoubi2009superradiant}. In order to consistently treat such an open system, one has to start from a master equation instead of the Schroedinger equation after having traced out the electromagnetic vacuum modes~\cite{gardiner2004quantum}. While the interaction between a pair of spins can be rather small at a larger distance, the collective effect of a sizable number of particles can still generate noticeable effects in this case\cite{zoubi2013excitons}.

Besides using polar molecules, which can possess relatively strong dipole moments~\cite{pollet2010supersolid}, another interesting implementation is based on using long lived atomic clock transitions in a differential light-shift-free magic wavelength lattice~\cite{takamoto2005optical,ludlow2008sr}, where one obtains extremely well controllable and precisely measurable systems to study even weak spin interactions~\cite{zhang2014spectroscopic} and collective decay via dipole-dipole energy exchange~\cite{zoubi2009superradiant}. For sufficient densities the particles' effective transition frequency and spontaneous decay is modified by dipole-dipole interaction~\cite{ostermann2012cascaded}, which in turn will influence the performance of a corresponding clock or super-radiant laser~\cite{maier2014superradiant}.

While the extremely small dipole moment of a clock transition keeps these interactions small, even tiny shifts and broadenings will ultimately influence clock accuracy and precision. Hence, reliable and converging numerical models are required to estimate these effects to many digits, in particular as one tries to work with as large as possible an ensemble to reduce measurement time and projection noise. For rather small atom numbers, up to about 10, a numerical solution of the full master equation is still possible~\cite{ostermann2012cascaded} and has showed that shifts and broadening can be non-negligible. For larger ensembles at low densities a so called cluster approach based on statistical averaging of important small particle number configurations has already produced first estimates of their scaling with the system's size~\cite{martin2013quantum}. Here, we focus on a more generally valid approach, namely the above mentioned mean-field plus pair-correlation method (MPC) to tackle large systems at high densities, i.e.~up to unit filling. The long range nature of the dipole coupling is accounted for by adding higher order corrections to the standard local factorization approach. In particular, for cavity mediated dipole interactions or coupling via nano-fibers even infinite range interactions have to be considered~\cite{gopalakrishnan2012frustration}. The focus of this work is put on developing the appropriate general numerical framework to treat such extended open spin lattices in various configurations and test their accuracy and convergence properties by means of the example of collective decay of a highly excited spin state. This should be the basis of future more specific work on concrete implementations of lattice clock Ramsey spectroscopy~\cite{ostermann2013protected} and super-radiant laser setups~\cite{bohnet2012steady,maier2014superradiant}.

This work is organized as follows. First, we give a short description of the system of coupled spins and introduce the corresponding master equation governing their time evolution including decay. Using generalized factorization assumptions for the density operator of the system we derive two approximate and numerically advantageous methods to calculate its time evolution. In the subsequent section we perform calculations for large ensembles studying the influence of long-range interactions. Here, we add an extensive numerical analysis to characterize the magnitude and scaling of the error of these two approximations depending on the geometry and the choice of initial state. Finally, we use this method to simulate systems of increasing size to study to which extend a finite sized sample can capture the dynamics of larger or even infinite systems.

\section{Interacting spin dynamics}
We consider a system consisting of $N$ two-level subsystems with transition frequency $\omega_0$ and decay rate $\gamma$ in an arbitrary spatial configuration. Each particle couples to the modes of the free electro-magnetic field and therefore all particles are indirectly coupled to one another. Mathematically, this problem can be simplified by treating the electro-magnetic modes as a single bath and introducing effective particle-particle interactions and effective decay of particle excitations into this bath, according to~\cite{ficek1986cooperative}.
The time evolution of the $N$ spins in a rotating frame corresponding to $\sum_i \omega_0 \sigma^z_i$ is then governed by a master equation
\begin{equation}
  \dot{\rho} = -\frac{i}{\hbar} \komm{H}{\rho} + \mathcal{L}[\rho]
  \label{eq:master-equation}
\end{equation}
with the Hamiltonian
\begin{equation}
  H = \sum_{ij;i \neq j} \hbar \Omega_{ij} \sigma_i^+ \sigma_j^-
  \label{eq:Hamiltonian}
\end{equation}
and Lindblad-term
\begin{equation}
  \mathcal{L}[\rho] = \frac{1}{2} \sum_{i,j} \Gamma_{ij}
                        (2\sigma_i^- \rho \sigma_j^+
                        - \sigma_i^+ \sigma_j^- \rho
                        - \rho \sigma_i^+ \sigma_j^-).
  \label{eq:Lindblad}
\end{equation}
The dipole-dipole interaction $\Omega_{ij} = \frac{3}{4} \gamma G(k_0 r_{ij})$ and the collective decay $\Gamma_{ij} = \frac{3}{2} \gamma F(k_0 r_{ij})$ can be obtained analytically with
\begin{subequations}
\begin{align}
  F(\xi) &= \alpha \frac{\sin \xi}{\xi}
            + \beta \left(
                  \frac{\cos \xi}{\xi^2} - \frac{\sin \xi}{\xi^3}
            \right)
  \\
  G(\xi) &= -\alpha \frac{\cos \xi}{\xi} + \beta \left(
                \frac{\sin \xi}{\xi^2} + \frac{\cos \xi}{\xi^3}
            \right)
\end{align}
\end{subequations}
with $\alpha = 1 -\cos^2 \theta$ and $\beta = 1-3 \cos^2 \theta$, where $\theta$ represents the angle between the line connecting atoms $i$ and $j$ and the common atomic dipole orientation~\cite{ficek1986cooperative}.

While in systems consisting of only very few particles we can study the time evolution by directly integrating the master equation, the exponential scaling of the dimension of the Hilbert space soon defeats any numerical abilities. To be able to represent the state of such a high particle number system in a computer one has to make simplifying assumptions about the form of the density matrix. In our following calculations we will truncate correlations between the particles at a certain order which greatly reduces the space needed to store the state of the system in memory and allows for treating larger particle numbers.

\subsection{Mean-field method: product state assumption}
In the first nontrivial approximation we neglect correlations altogether and assume that the system is at all times in a product state of the subsystems at each site. The density matrix is approximated by $\rho = \bigotimes_k \rho^{(k)}$, which is also called {\sl mean-field approximation}. The time evolution of the system is then governed by the local on site density matrices, which for two-level systems can be obtained from a complete set of expectation values for each spin, i.e.~the expectation values of the Pauli operators $\expect{\sigma_x}$, $\expect{\sigma_y}$ and $\expect{\sigma_z}$ for a spin $1/2$ system. Using this Pauli representation we need three real numbers to characterize the state of each of the two-level sub-systems at a certain point in time. The resulting equations for the local spin provide an intuitive insight into the corresponding physics. Explicitly we get:
\begin{subequations}
\label{eq:mf-timeevolution}
\begin{align}
  \expect{\dot{\sigma_k^x}}  &=
      \sum_{i;i \neq k} \Omega_{ki} \expect{\sigma_i^y\sigma_k^z}
      -\frac{1}{2} \gamma \expect{\sigma_k^x}
      -\frac{1}{2} \sum_{i;i \neq k} \Gamma_{ki} \expect{\sigma_i^x\sigma_k^z}
\\
  \expect{\dot{\sigma_k^y}}  &=
      -\sum_{i;i \neq k} \Omega_{ki} \expect{\sigma_i^x\sigma_k^z}
      -\frac{1}{2} \gamma \expect{\sigma_k^y}
      -\frac{1}{2} \sum_{i;i \neq k} \Gamma_{ki} \expect{\sigma_i^y\sigma_k^z}
\\
  \expect{\dot{\sigma_k^z}}  &=
        -i \sum_{i;i \neq k} \Omega_{ki} \Big(\expect{\sigma_k^x\sigma_i^y} - \expect{\sigma_i^x\sigma_k^y}\Big)
        +\gamma \big(1 - \expect{\sigma_k^z}\big)
        \nonumber\\&\qquad
        +\frac{1}{2} \sum_{i;i \neq k} \Gamma_{ki} \Big(\expect{\sigma_k^x\sigma_i^x} + \expect{\sigma_i^y\sigma_k^y}\Big)
\end{align}
\end{subequations}
These equations still contain two-particle expectation values of the form $\expect{\sigma_i^\alpha\sigma_j^\beta}$, which according to our above assumption can be factorized, i.e.~$\expect{\sigma_i^\alpha\sigma_j^\beta} \approx \expect{\sigma_i^\alpha}\expect{\sigma_j^\beta}$. As we will see in the next section for weak inter-particle interactions this gives a surprisingly good approximation to the interaction induced shifts and can also account for spatial inhomogeneities of the system.

\subsection{Extended mean-field method including pair-correlations (MPC)}
As a next-order correction to the above mean-field approach we now include pair-correlations but still neglect all higher-order correlations. To this end the density matrix can be approximated by $\rho = \bigotimes_i \rho^{(i)} + \sum_{j<k}\Big(\rho^{(j,k)}\otimes \bigotimes_{i \neq j,k} \rho^{(i)}\Big)$, where the first term is the previously used product state and the correlations are captured in the operators $\rho^{(j,k)}$. The correlations thus have to be chosen to generate vanishing single particle expectation values, i.e.~$\Tr{\sigma_i^\alpha \rho^{(j,k)}} = 0$. Deriving the equations of motion in terms of expectation values of Pauli operators leads to the same equations as in the mean-field case (eq.~\ref{eq:mf-timeevolution}). The two-particle expectation values are then determined via a set of additional equations for the expectation values of all two-particle Pauli operator pairs of the type $\expect{\sigma_i^\alpha \sigma_j^\beta}$. In principle, there are nine such quantities for any pair of particles $\rho^{(j,k)}$. For symmetry reasons three of them are trivially obtained from the others. Similarly to the mean-field in the equations for these two-particle correlations higher order three-particle correlations appear, which based on our assumption of the form of the density operator are again approximated by
\begin{align}
  \expect{\sigma_i^\alpha\sigma_j^\beta\sigma_k^\gamma} &\approx
      -2\expect{\sigma_i^\alpha}\expect{\sigma_j^\beta}\expect{\sigma_k^\gamma}
      + \expect{\sigma_i^\alpha}\expect{\sigma_j^\beta\sigma_k^\gamma}
      + \expect{\sigma_j^\beta}\expect{\sigma_i^\alpha\sigma_k^\gamma}
      \nonumber\\&\qquad
      + \expect{\sigma_k^\gamma}\expect{\sigma_i^\alpha\sigma_j^\beta}.
\end{align}
Although the resulting equations of motions for the two-particle correlations are arguably bulky, we want to display them explicitly, as the form an essential basis of our work.
\begin{subequations}
\label{eq:mpc-timeevolution}
\begin{align}
  \expect{\dot{\sigma_k^x\sigma_l^x}} &=
      \sum_{j;j \neq k,l} \Omega_{kj} \expect{\sigma_k^z\sigma_l^x\sigma_j^y}
       + \sum_{j;j \neq k,l} \Omega_{lj} \expect{\sigma_k^x\sigma_l^z\sigma_j^y}
    \nonumber\\&\qquad
      - \gamma \expect{\sigma_k^x\sigma_l^x}
      + \Gamma_{kl} \Big(
              \expect{\sigma_k^z\sigma_l^z}
              - \frac{1}{2} \expect{\sigma_k^z}
              - \frac{1}{2} \expect{\sigma_l^z}
        \Big)
    \nonumber\\&\quad
        - \frac{1}{2} \sum_{j;j \neq k,l} \Gamma_{kj}
              \expect{\sigma_k^z\sigma_l^x\sigma_j^x}
        - \frac{1}{2} \sum_{j;j \neq k,l} \Gamma_{lj}
              \expect{\sigma_k^x\sigma_l^z\sigma_j^x}
\\
  \expect{\dot{\sigma_k^y\sigma_l^y}}
  &= - \sum_{j;j \neq k,l} \Omega_{kj}
          \expect{\sigma_k^z\sigma_l^y\sigma_j^x}
        - \sum_{j;j \neq k,l} \Omega_{lj}
          \expect{\sigma_k^y\sigma_l^z\sigma_j^x}
    \nonumber\\&\qquad
        - \gamma \expect{\sigma_k^y\sigma_l^y}
        + \Gamma_{kl}\Big(
              \expect{\sigma_k^z\sigma_l^z}
            -\frac{1}{2} \expect{\sigma_k^z}
            -\frac{1}{2} \expect{\sigma_l^z}
        \Big)
    \nonumber\\&\quad
        -\frac{1}{2} \sum_{j;j \neq k,l} \Gamma_{kj}
              \expect{\sigma_k^z\sigma_l^y\sigma_j^y}
        -\frac{1}{2} \sum_{j;j \neq k,l} \Gamma_{lj}
              \expect{\sigma_k^y\sigma_l^z\sigma_j^y}
\\
  \expect{\dot{\sigma_k^z\sigma_l^z}}
    &= \sum_{j;j \neq k,l} \Omega_{kj} \Big(
          \expect{\sigma_k^y\sigma_l^z\sigma_j^x}
          - \expect{\sigma_k^x\sigma_l^z\sigma_j^y}
        \Big)
    \nonumber\\&\qquad
        +\sum_{j;j \neq k,l} \Omega_{lj} \Big(
          \expect{\sigma_k^z\sigma_l^y\sigma_j^x}
          -\expect{\sigma_k^z\sigma_l^x\sigma_j^y}
        \Big)
    \nonumber\\&\quad
        - 2 \gamma \expect{\sigma_k^z\sigma_l^z}
        + \gamma \big(\expect{\sigma_l^z} + \expect{\sigma_k^z}\big)
    \nonumber\\&\quad
        +\Gamma_{kl}\Big(
              \expect{\sigma_k^y\sigma_l^y}
              + \expect{\sigma_k^x\sigma_l^x}
        \Big)
    \nonumber\\&\quad
        +\frac{1}{2} \sum_{j;j \neq k,l} \Gamma_{kj} \Big(
              \expect{\sigma_k^x\sigma_l^z\sigma_j^x}
              +\expect{\sigma_k^y\sigma_l^z\sigma_j^y}
        \Big)
    \nonumber\\&\qquad
        +\frac{1}{2} \sum_{j;j \neq k,l} \Gamma_{lj} \Big(
              \expect{\sigma_k^z\sigma_l^x\sigma_j^x}
              +\expect{\sigma_k^z\sigma_l^y\sigma_j^y}
        \Big)
\\
  \expect{\dot{\sigma_k^x\sigma_l^y}}
    &= \Omega_{kl}\Big(
          \expect{\sigma_k^z}
          - \expect{\sigma_l^z}
        \Big)
        +\sum_{j;j \neq k,l} \Omega_{kj}
          \expect{\sigma_k^z\sigma_l^y\sigma_j^y}
    \nonumber\\&\qquad
        -\sum_{j;j \neq k,l} \Omega_{lj}
          \expect{\sigma_k^x\sigma_l^z\sigma_j^x}
        - \gamma \expect{\sigma_k^x\sigma_l^y}
    \nonumber\\&\quad
        - \frac{1}{2} \sum_{j;j \neq k,l} \Gamma_{kj}
              \expect{\sigma_k^z\sigma_l^y\sigma_j^x}
        - \frac{1}{2} \sum_{j;j \neq k,l} \Gamma_{lj}
              \expect{\sigma_k^x\sigma_l^z\sigma_j^y}
\\
  \expect{\dot{\sigma_k^x\sigma_l^z}}
    &= \Omega_{kl}
          \expect{\sigma_l^y}
        +\sum_{j;j \neq k,l} \Omega_{kj}
          \expect{\sigma_k^z\sigma_l^z\sigma_j^y}
    \nonumber\\&\quad
        +\sum_{j;j \neq k,l} \Omega_{lj} \Big(
          \expect{\sigma_k^x\sigma_l^y\sigma_j^x}
          -\expect{\sigma_k^x\sigma_l^x\sigma_j^y}
        \Big)
  \nonumber\\&\quad
  - \frac{3}{2} \gamma \expect{\sigma_k^x\sigma_l^z}
      + \gamma \expect{\sigma_k^x}
      - \Gamma_{kl}\Big(
            \expect{\sigma_k^z\sigma_l^x}
            -\frac{1}{2} \expect{\sigma_l^x}
        \Big)
    \nonumber\\&\quad
        - \frac{1}{2} \sum_{j;j \neq k,l} \Gamma_{kj}
              \expect{\sigma_k^z\sigma_l^z\sigma_j^x}
    \nonumber\\&\quad
        + \frac{1}{2} \sum_{j;j \neq k,l} \Gamma_{lj} \Big(
              \expect{\sigma_k^x\sigma_l^x\sigma_j^x}
              +\expect{\sigma_k^x\sigma_l^y\sigma_j^y}
        \Big)
\\
  \expect{\dot{\sigma_k^y\sigma_l^z}}
    &= -\Omega_{kl} \expect{\sigma_l^x}
        -\sum_{j;j \neq k,l} \Omega_{kj}
          \expect{\sigma_k^z\sigma_l^z\sigma_j^x}
    \nonumber\\&\quad
        +\sum_{j;j \neq k,l} \Omega_{lj} \Big(
          \expect{\sigma_k^y\sigma_l^y\sigma_j^x}
          -\expect{\sigma_k^y\sigma_l^x\sigma_j^y}
        \Big)
    \nonumber\\&\quad
      - \frac{3}{2} \gamma \expect{\sigma_k^y\sigma_l^z}
      + \gamma \expect{\sigma_k^y}
      - \Gamma_{kl}\Big(
              \expect{\sigma_k^z\sigma_l^y}
            - \frac{1}{2}\expect{\sigma_l^y}
        \Big)
    \nonumber\\&\quad
        - \frac{1}{2} \sum_{j;j \neq k,l} \Gamma_{kj}
              \expect{\sigma_k^z\sigma_l^z\sigma_j^y}
    \nonumber\\&\quad
        + \frac{1}{2} \sum_{j;j \neq k,l} \Gamma_{lj} \Big(
              \expect{\sigma_k^y\sigma_l^x\sigma_j^x}
              +\expect{\sigma_k^y\sigma_l^y\sigma_j^y}
        \Big)
\end{align}
\end{subequations}
Note, that the number of equations to be solved increases quadratically with the number of particles, as we include all possible two-particle combinations. This is exponentially slower than the growth of the corresponding Hilbert space. In many cases one might even be able to restrict this to nearest neighbor couplings only, but for long range dipole or cavity mediated interactions, in which we are interested here, no such truncations can be performed safely. In principle the method, which in many respects resembles the known cumulant expansion method~\cite{henschel2010cavity}, can be extended towards higher order. However, as we will see below, it is already very accurate for our purposes so that we will not pursue this task.

\section{Numerical accuracy of the mean-field method and second order corrections}
In the previous section we have presented two numerical approaches to approximate the master equation (eq.~\ref{eq:master-equation}) by neglecting higher-order quantum correlations. In order to examine for which conditions these assumptions lead to accurate solutions, we compare these approximations with the numerical solution of the full master equation for different spatial arrangements, numbers of particles and initial states. Additionally we also calculate the case of independent particles, which allows us to identify examples where the error of the approximation is small due to a negligible influence of the dipole-dipole-interaction and the collective decay.

\subsubsection{Spin dynamics}
\begin{figure*}[ht]
  \includegraphics{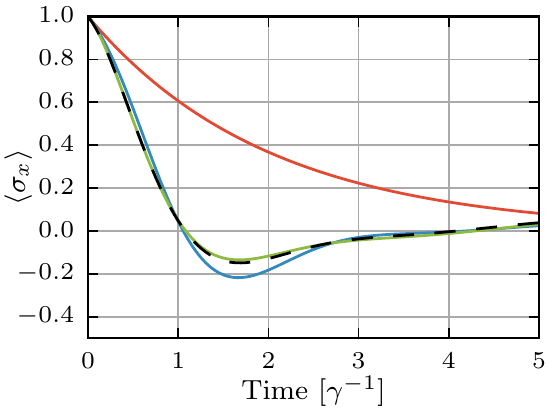}
  \includegraphics{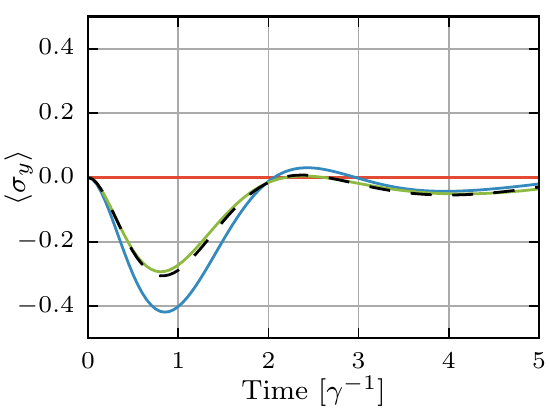}
  \includegraphics{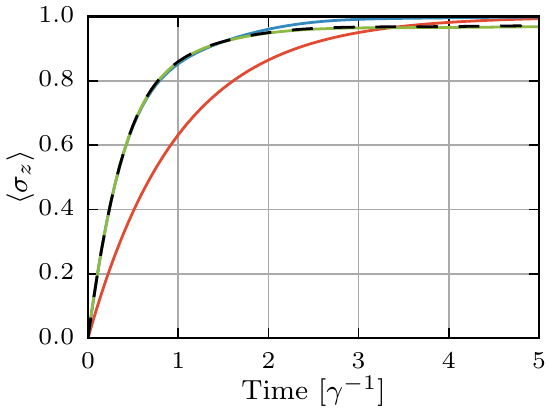}
  \caption{Time evolution of the expectation values of the Pauli operators $\sigma_x$, $\sigma_y$ and $\sigma_z$ of the central spin in a chain consisting of 7 spins with spin-spin distance $d=0.15 \lambda_0$. The system is simulated using independent spins (red), the mean-field method (blue), MPC (green) and by solving the whole master equation (dashed black). The dipole is orientated orthogonally to the chain.}
  \label{fig:timeevolution-chain}
\end{figure*}

\begin{figure*}[ht]
  \includegraphics{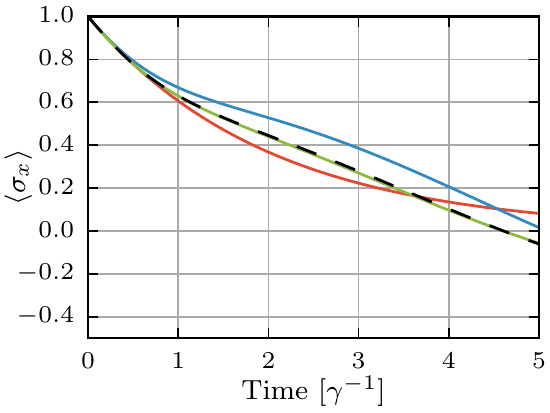}
  \includegraphics{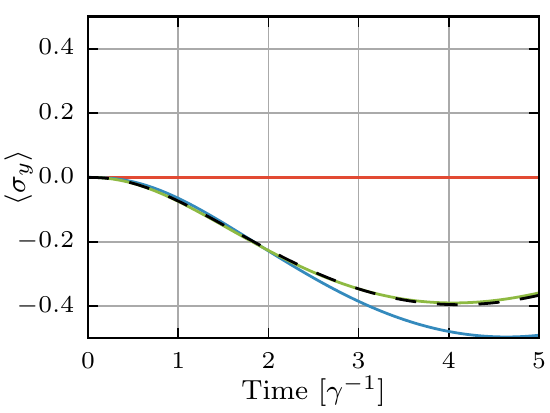}
  \includegraphics{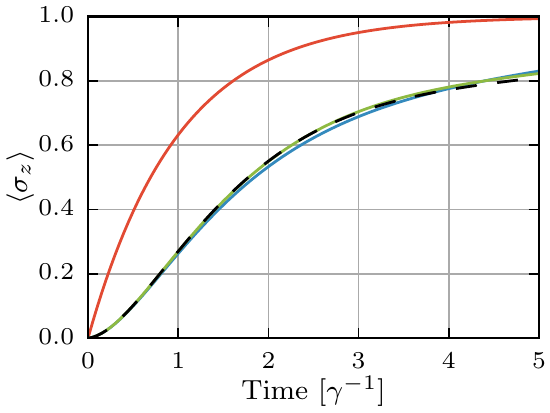}
  \caption{Time evolution of the expectation values of the Pauli operators $\sigma_x$, $\sigma_y$ and $\sigma_z$ of the central spin in a 2D-square lattice consisting of 3x3 spins with nearest spin-spin distance $d=0.5 \lambda_0$. The system is simulated using independent spins (red), the mean-field method (blue), MPC (green) and by solving the whole master equation (dashed black). The dipole is orientated orthogonally to the plane.}
  \label{fig:timeevolution-squarelattice}
\end{figure*}

\begin{figure*}[ht]
  \includegraphics{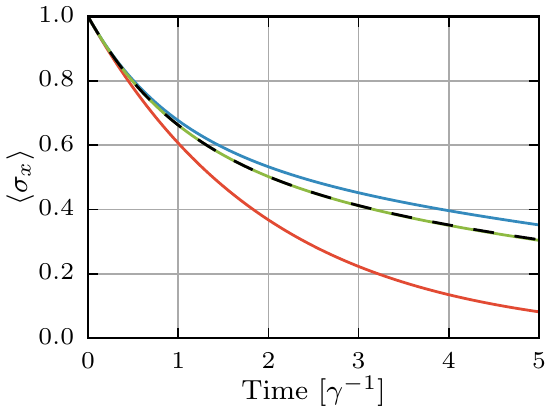}
  \includegraphics{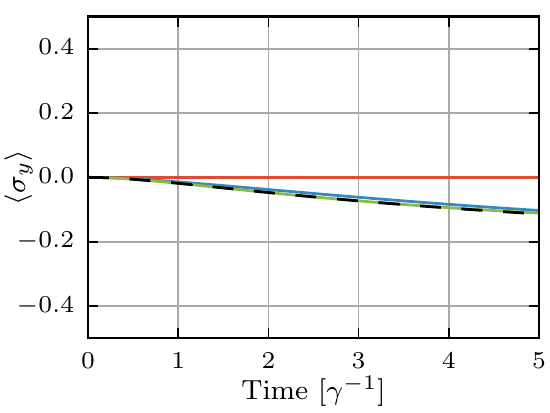}
  \includegraphics{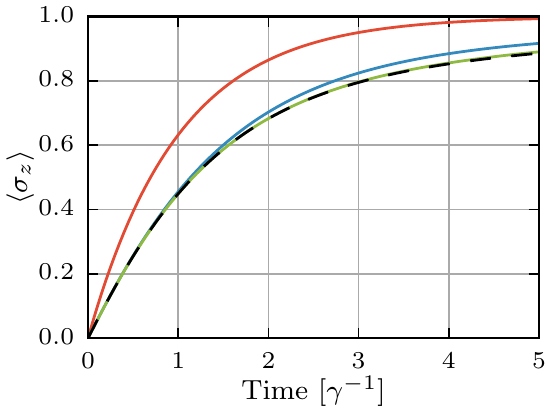}
  \caption{Time evolution of the expectation values of the Pauli operators $\sigma_x$, $\sigma_y$ and $\sigma_z$ of a single spin in a cube configuration with nearest spin-spin distance $d=0.6 \lambda_0$. The system is simulated using independent spins (red), the mean-field method (blue), MPC (green) and by solving the whole master equation (dashed black). The dipole is orientated orthogonally to an arbitrary face of the cube.}
  \label{fig:timeevolution-cube}
\end{figure*}

To obtain a first intuitive understanding for the quality of the different methods we compare the time evolution of the expectation values of the Pauli operators for three different geometries, i.e.~a chain (fig.~\ref{fig:timeevolution-chain}), a square lattice (fig.~\ref{fig:timeevolution-squarelattice}) and a cube (fig.~\ref{fig:timeevolution-cube}).
As a generic physical example, we start with a product state of all spins pointing in x-direction. This is the state prepared in the first step of a typical Ramsey spectroscopy procedure. It is fully superradiant when all particles are confined in a very small spatial volume.
Clearly, the dynamics of all three cases is significantly different, but they all share certain features. First, the solution of the full master equation deviates drastically from the independent particle case, which means that the effect of the collective interaction is significant. This deviation is almost perfectly captured by the second order MPC solution, which is, at least visually, almost identical to the full solution of the master equation. Surprisingly, the mean-field solution shows a qualitatively similar behavior already, although it is noticeably not as accurate. Note, that for the case of the cube (fig.~\ref{fig:timeevolution-cube}) both methods predict the subradiance of the initial spin state well~\cite{ostermann2013protected}. Let us now turn from a visual to a more systematic numerical error estimation.

\subsubsection{Systematic accuracy analysis}
\begin{figure*}[ht]
  \includegraphics{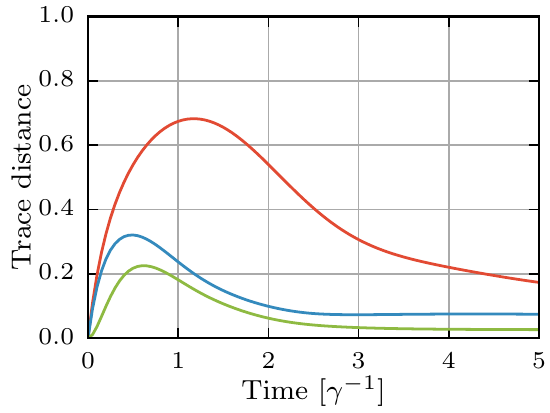}
  \includegraphics{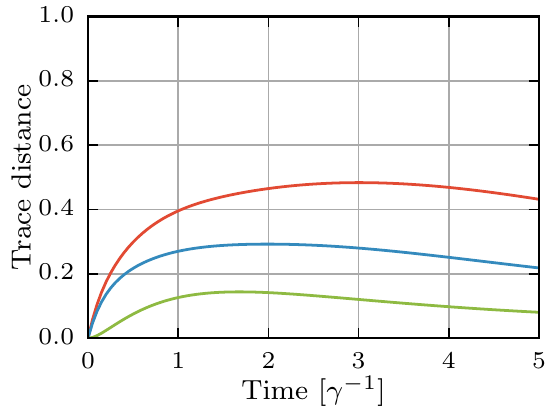}
  \includegraphics{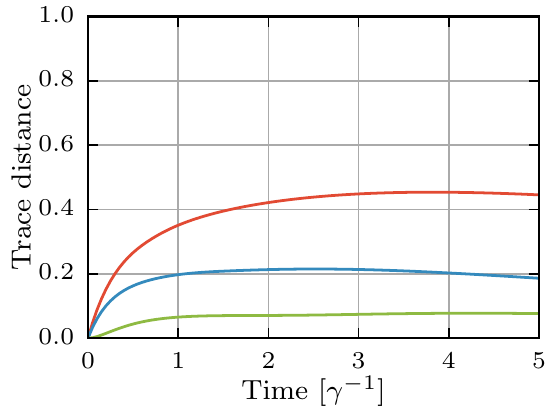}
  \caption{Trace distance between the density operators calculated by the master equation and density operators calculated for independent spins (red), mean-field (blue) and MPC (green) for the previously discussed chain configuration (a), 3x3 square lattice (b) and cube configuration (c).}
  \label{fig:timeevolution-tracedistance}
\end{figure*}

In the following we will perform a more rigorous, quantitative analysis for a large range of parameters. In order to do this effectively we need a simple measure of accuracy of the different methods. A frequently used tool, especially in quantum information, is the trace distance which is defined as $\mathcal{T}(\rho, \sigma)=\frac{1}{2}|\lambda_i|$ where the $\lambda_i$ are the eigenvalues of the matrix representation of $\rho-\sigma$. For qubits this measure has a very intuitive interpretation, it is just half of the geometric distance of the two states on the Bloch sphere. In fig.~\ref{fig:timeevolution-tracedistance} we use this trace distance between the solution of the master equation and the previously presented numerical methods at equal points in time to characterize the error of the different approximations.
In all our examples we initially start in a product state, which means that the error at $t=0$ is always zero and since no additional pumping is included the system decays to the ground state and the trace distance in the long time limit will vanish for all numerical methods. Instead of inspecting the variation of the trace distance over time we will use the time-maximum of the trace distance as a characterization of the error.

\subsection{Geometry dependence}
\begin{figure*}[ht]
  \includegraphics{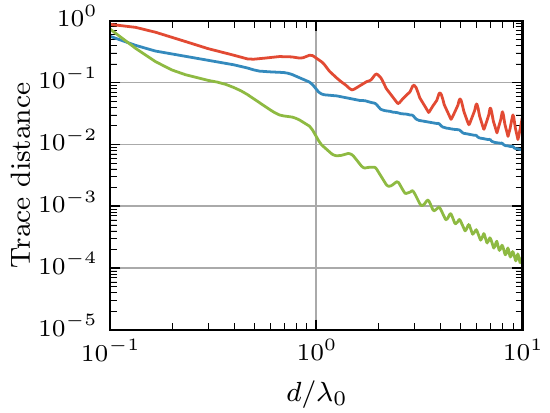}
  \includegraphics{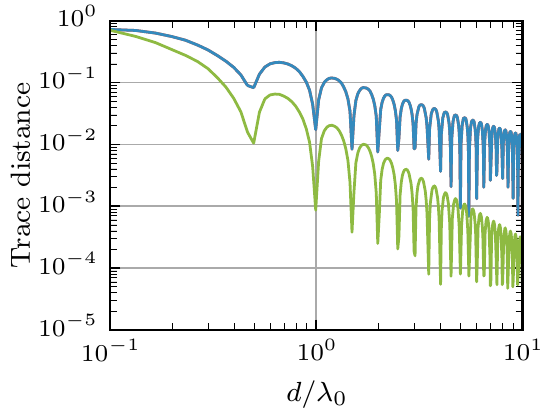}
  \includegraphics{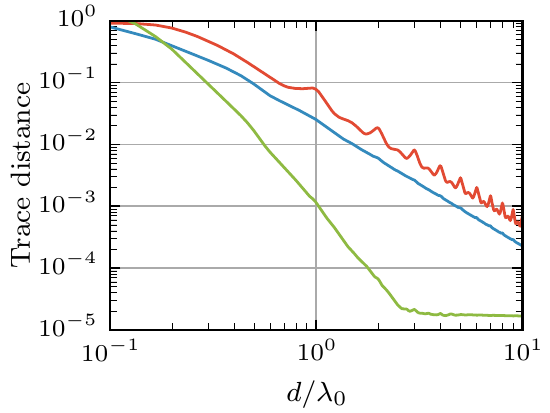}
  \caption{Distance dependency of the time-maximum of the trace distance between the results of the master equation and the results of an independent evolution (red), mean-field (blue) and MPC (green) for a chain consisting of 8 spins for different initial states and dipole orientations. (a) $\Theta=\pi/2$, $e_{dipole}=e_z$. (b) $\Theta=\pi$, $e_{dipole}=e_z$. (c) $\Theta=\pi/2$, $e_{dipole}=e_x$.}
  \label{fig:erroranalysis-geometry-chain}
\end{figure*}

\begin{figure*}[ht]
  \includegraphics{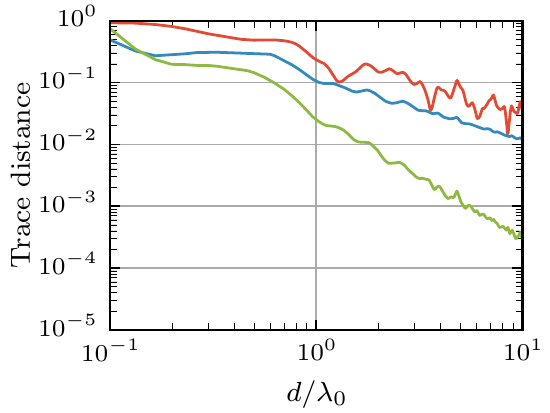}
  \includegraphics{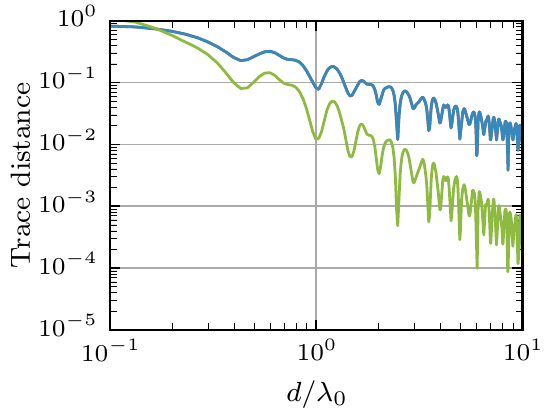}
  \includegraphics{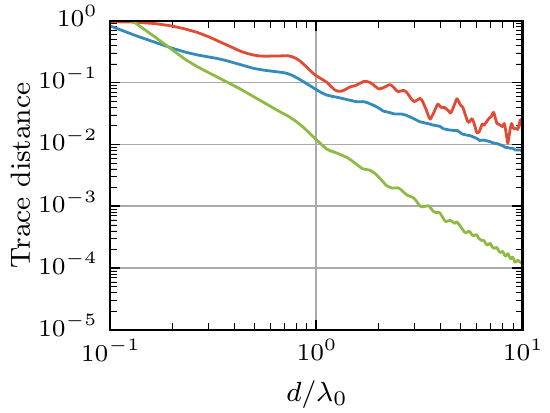}
  \caption{Distance dependency of the time-maximum of the trace distance between results of the master equation and results of independent evolution (red), mean-field (blue) and MPC (green) for a square lattice consisting of 3x3 spins for different initial states and dipole orientations. (a) $\Theta=\pi/2$, $e_{dipole}=e_z$. (b) $\Theta=\pi$, $e_{dipole}=e_z$. (c) $\Theta=\pi/2$, $e_{dipole}=e_x$.}
  \label{fig:erroranalysis-geometry-squarelattice}
\end{figure*}

\begin{figure*}[ht]
  \includegraphics{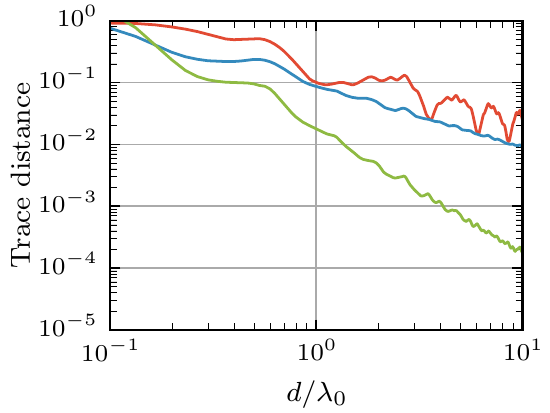}
  \includegraphics{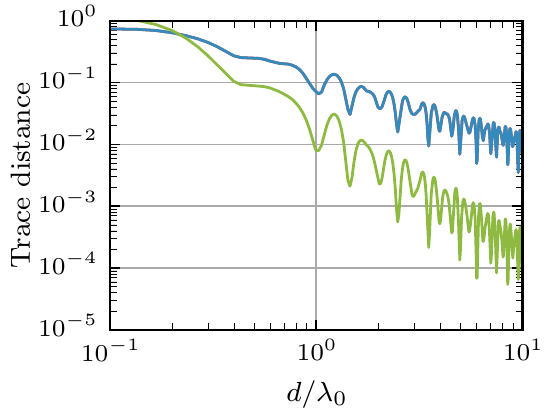}
  \includegraphics{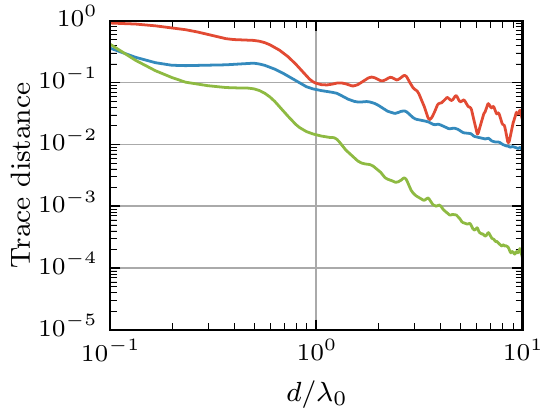}
  \caption{Distance dependency of the time-maximum of the trace distance between the results of the master equation and the results of independent evolution (red), mean-field (blue) and MPC (green) for 8 spins in a cube configuration for different initial states and dipole orientations. (a) $\Theta=\pi/2$, $e_{dipole}=e_z$. (b) $\Theta=\pi$, $e_{dipole}=e_z$. (c) $\Theta=\pi/2$, $e_{dipole}=\frac{1}{\sqrt{3}}(e_x+e_y+e_z)$.}
  \label{fig:erroranalysis-geometry-cube}
\end{figure*}

In this section we study the geometry dependence of the error of the numerical methods measured by the previously introduced time maximum of the trace distance. We distinguish between systems of different dimensionality, a 1D chain consisting of 8 particles (fig.~\ref{fig:erroranalysis-geometry-chain}), a 3x3 section of a 2D square lattice (fig.~\ref{fig:erroranalysis-geometry-squarelattice}) and a cube as a 3D configuration (fig.~\ref{fig:erroranalysis-geometry-cube}). For each of these examples we calculate the dependence of the error on the distance between the particles. Further on we vary the initial state and the orientation of the polarization vector and show three typical results. In the following the initial state is characterized by $\Theta$ which measures the polar angle between a given state towards the ground state on the Bloch sphere.
Several interesting features stand out immediately. The bigger the distance between the particles, the smaller the error of neglecting higher-order correlations. As can be seen from the trace distance between the solution of the master equation and the independently decaying case this is to some degree an artifact of the decreasing strength of the dipole-dipole interaction which in the far field limit has a $\frac{1}{r}$ dependency but at least for MPC the error decreases much faster. In nearly all cases the mean-field approach yields a noticeable improvement, yet, when all spins start in the excited state it reproduces the results of independent particles only. In fact, as one can show from the mean-field equations, in this case the time evolution is completely identical, so mean-field results in no improvement over simply ignoring the collective effects.

\subsection{Initial state dependence}
\begin{figure*}[ht]
  \includegraphics{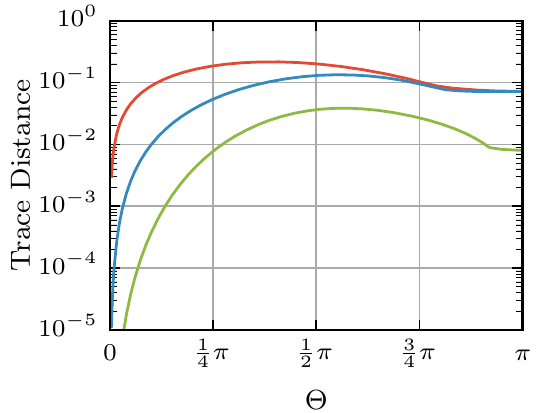}
  \includegraphics{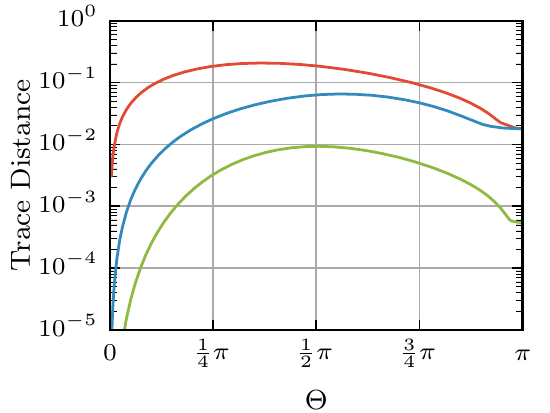}
  \includegraphics{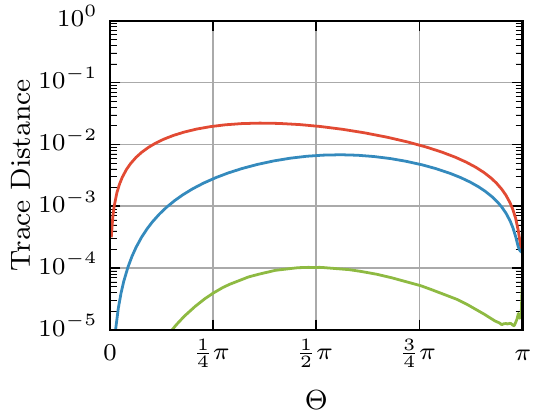}
  \caption{Dependence of the time-maximum of the trace distance between the results of the master equation and the results of an independent evolution (red), mean-field (blue) and MPC(green) on the initial Bloch state characterized by the polar angle $\Theta$ for a chain consisting of 6 spins with spin-spin distance $d=0.5 \lambda_0$ (a), $d=1.0 \lambda_0$(b) and $d=10.0 \lambda_0$ (c).}
  \label{fig:erroranalysis-initialstate}
\end{figure*}

To further analyze the dependence of the error on the initial state we consider a chain of six particles with three different particle distances. Initially the system is in a product state where all single particles are in the same Bloch state. For simplicity we only consider pure states and since the time evolution is invariant under a global rotation around the z-axis the only remaining variable is the polar angle $\Theta$. In fig.~\ref{fig:erroranalysis-initialstate} the dependence of the error on this polar angle is shown. For $\Theta=0$ the system is in the ground state and the error vanishes. For a small excitation the mean-field method gives a substantial improvement compared to the independently decaying system but for a nearly totally excited state the advantage disappears more and more. In contrast, MPC performs convincingly for all initial states.

\subsection{Spin-number dependence}
\begin{figure*}[ht]
  \includegraphics{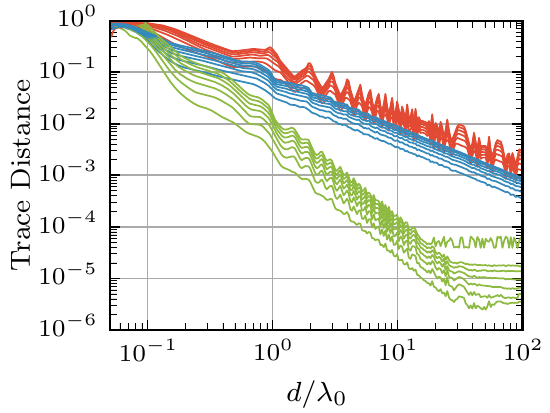}
  \includegraphics{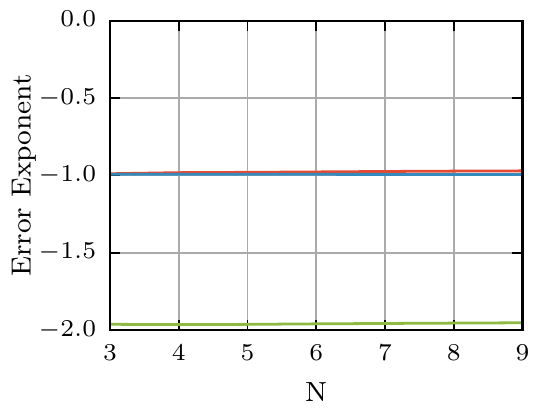}
  \includegraphics{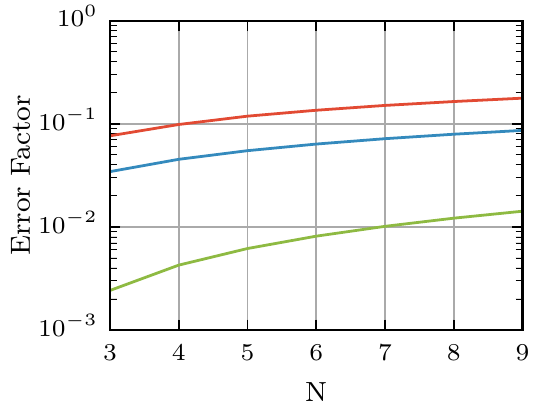}
  \caption{Dependence analysis of the time-maximum trace distance between results of the master equation and results of an independent evolution (red), mean-field (blue) and MPC (green) on the spin distance of a chain consisting of $N=3\hdots9$ spins. Higher spin numbers correspond to slightly increased trace distances (a). An approximation of the trace distances by $C_N*d^{k_N}$ results in the spin-number dependency of the error-exponent $k_N$ (b) and the error-factor $C_N$ (c).}
  \label{fig:erroranalysis-spinnumber}
\end{figure*}

Finally, we investigate the dependence of the error on the number of particles in the system, i.e.~a chain consisting of $N$ particles. The result of this analysis is shown in fig.~\ref{fig:erroranalysis-spinnumber}a. In this double logarithmic plot the error appears to be nearly linear but slightly shifted for varying particle numbers which leads us to the following estimate for the error
\begin{align}
  \mathrm{err}(N,d) = C_N*d^{k_N}.
\end{align}
The exponent $k_N$ and the factor $C_N$ can be extracted from this error plot and are shown in fig.~\ref{fig:erroranalysis-spinnumber}b and fig.~\ref{fig:erroranalysis-spinnumber}c respectively. The error exponent turns out to be independent of the number of particles and is -1 for independently decaying spins which is not surprising since the collective interaction in the far field drops with $\frac{1}{r}$. However, increasing the distances doesn't improve the mean-field results whereas MPC has an error exponent of -2 and gains drastically on accuracy.

\section{Approximation of very large (infinite) systems}
Recent research on the effect of geometry on the perturbation of the spin dynamics by collective interactions was mostly limited to systems consisting of only very few atoms. In lack of better alternatives one might be tempted to extrapolate results obtained from these small-sized systems to larger ensembles but in general this attempt could fail miserably. Armed with the knowledge about the accuracy of the mean-field and MPC methods and their ability to simulate moderately large systems we can use them to investigate how many particles are needed to make satisfying statements about infinite systems. More precisely, we want to know how collective spin quantities of the type $\frac{1}{N} \sum_i \expect{\sigma^\alpha_i}$ will change for different numbers of particles. Of course, it is not a priori clear if these expectation values will converge at all. To answer this question we will study two different examples.

\subsection{Linear equidistant chain}
\begin{figure*}[ht]
  \includegraphics{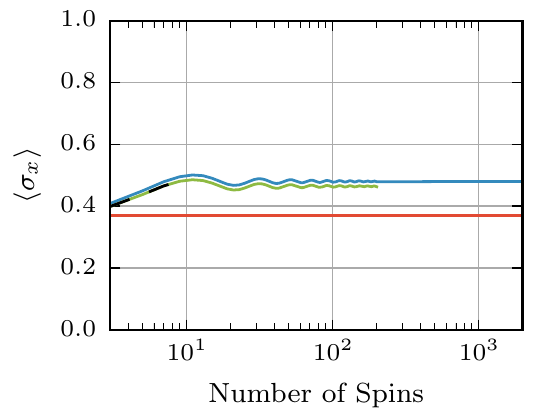}
  \includegraphics{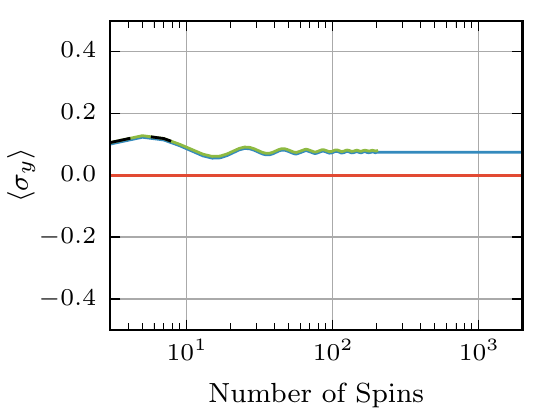}
  \includegraphics{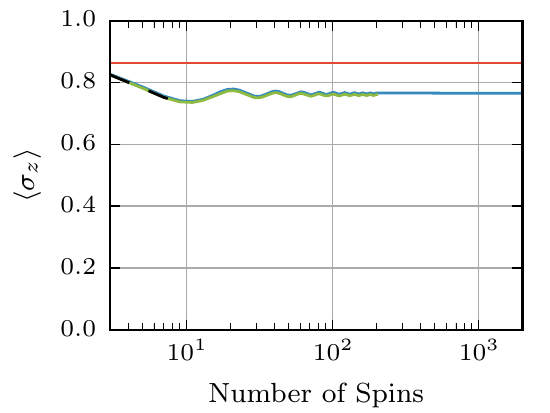}
  \caption{Expectation values of the Pauli operators $\expect{\sigma_x}$, $\expect{\sigma_y}$ and $\expect{\sigma_z}$ of the central spin in a chain consisting of $N$ spins with distance $d=0.9\lambda_0$ after a time evolution for $2\gamma^{-1}$. Initially all spins are in the state $\expect{\sigma_x}=1$ and the system is solved for independent spins (red), using mean-field (blue), MPC (green) and the master-equation (dashed black).}
  \label{fig:timeevolution-centralspin-chain}
\end{figure*}

\begin{figure*}[ht]
  \includegraphics{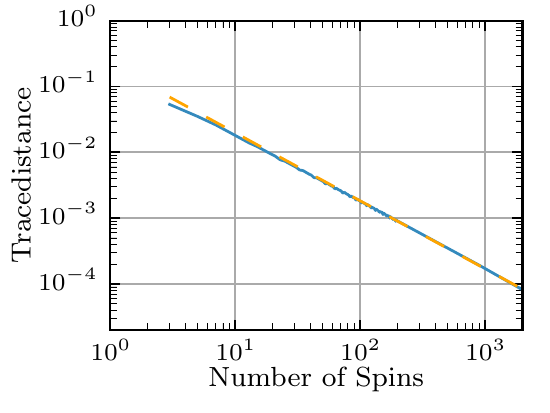}
  \includegraphics{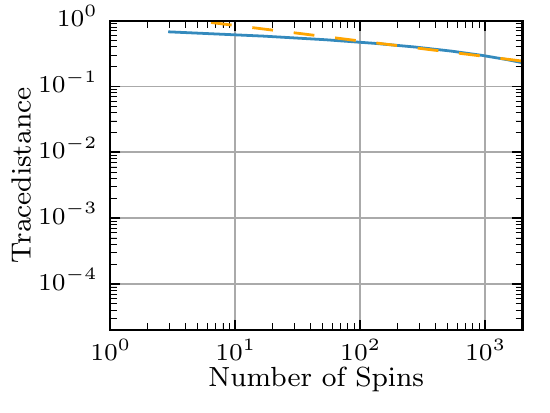}
  \includegraphics{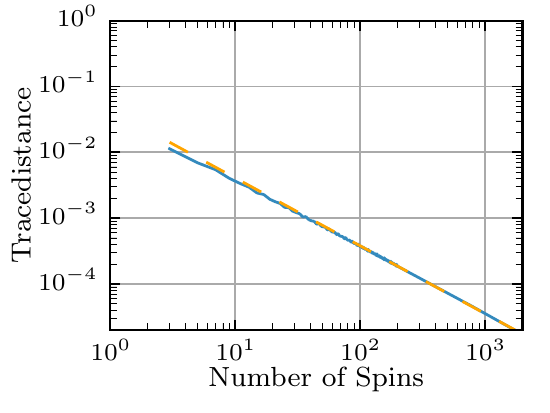}
  \caption{Time-maximum of the trace-distance between the reduced density matrix of the central spin state in a chain consisting of $N$ spins compared to the chain consisting of $N_\mathrm{max}^\mathrm{meanfield}=20001$ (blue) where both quantities are results of mean-field simulations. Approximation of this trace-distance by $C_d*N^{k_d}$ (dashed yellow) for different spin-spin distances $d=0.7 \lambda_0$(a), $d=1.0 \lambda_0$(b) and $d=3.7 \lambda_0$(c)}
  \label{fig:centralsipn-chain-meanfield}
\end{figure*}

\begin{figure*}[ht]
  \includegraphics{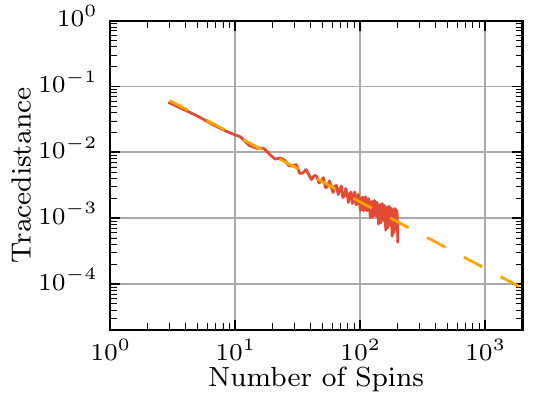}
  \includegraphics{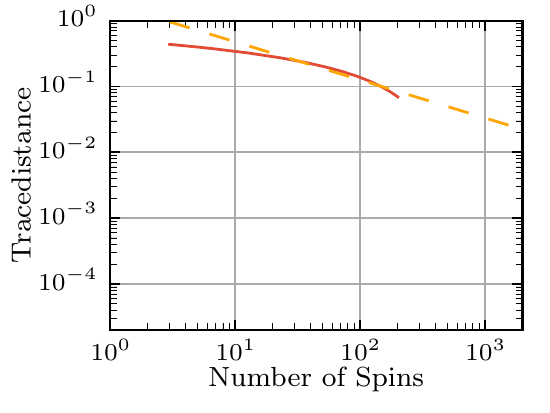}
  \includegraphics{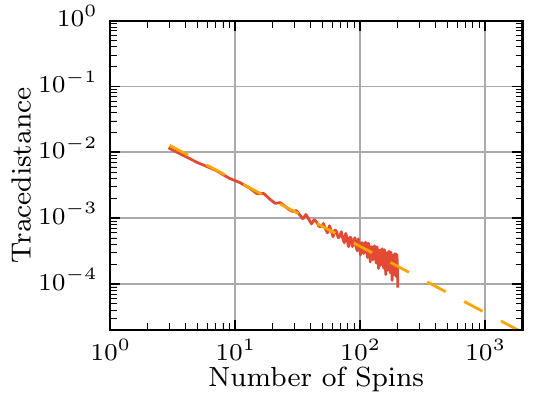}
  \caption{Time-maximum of the trace-distance between the reduced density matrix of the central spin in a chain consisting of $N$ spins compared to the chain consisting of $N_\mathrm{max}^\mathrm{MPC}=401$ (blue) where both quantities are results of MPC simulations for different spin-spin distances $d=0.7 \lambda_0$(a), $d=1.0 \lambda_0$(b) and $d=3.7 \lambda_0$(c). Approximation of this trace-distance by $C_d*N^{k_d}$ (dashed yellow).}
  \label{fig:centralsipn-chain-mpc}
\end{figure*}

\begin{figure*}[ht]
  \includegraphics{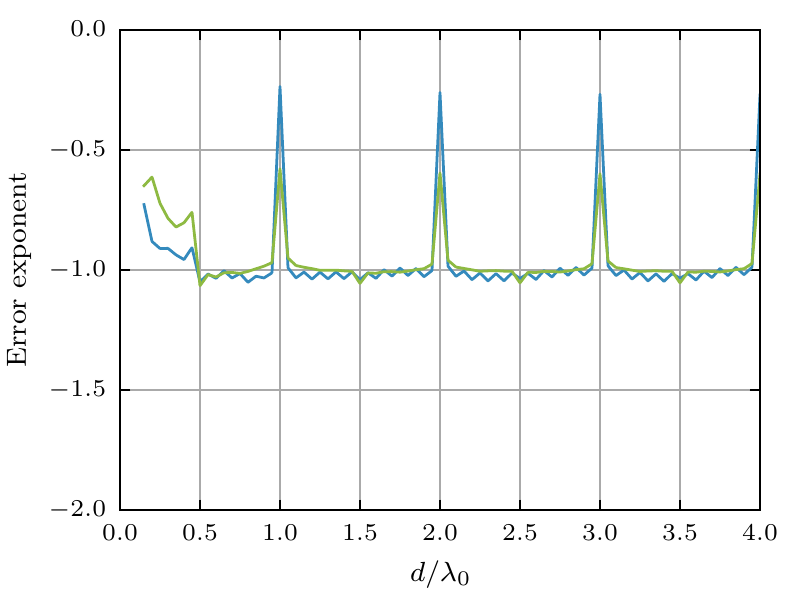}
  \includegraphics{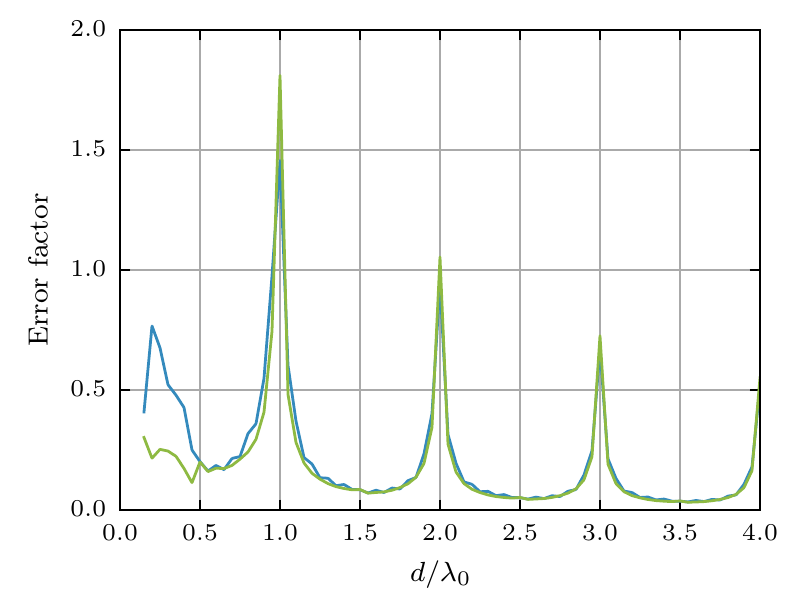}
  \caption{Error exponent $k_d$ and Error factor $C_d$ dependence on the spin-spin distance determined in the previous fits $C_d*N^{k_d}$ using mean-field (blue) and MPC (green) in the case of a chain.}
  \label{fig:centralsipn-chain-tdestimate}
\end{figure*}

We consider a $N$-particle spin chain with particle distance $d$ and calculate the dynamics of the whole system where initially all spins are in the $\expect{\sigma_x}=1$ state. Tracing out all but the innermost spin allows us to compare the dynamics of this single spin for a varying number of surrounding particles. The result of this analysis for a certain distance $d$ after an integration time of $2\gamma^{-1}$ is shown in fig.~\ref{fig:timeevolution-centralspin-chain}. Fortunately, all methods yield more or less the same result and differ significantly from the independently decaying case, indicating that the variation for small particle numbers and the ultimate convergence for large systems is not a numerical artifact.
This result hints at the fact that a suitable number of particles will indeed give a usable approximation of big systems. To consolidate this claim we perform a more extensive and quantitative test. What we actually would like to test is how much the time evolution of the single central spin in a chain consisting of $N$ particles differs from the time evolution of a spin in an infinite chain. However, we are not aware of a method to solve the infinite chain exactly which leaves us with the option to compare the central spin of a $N$ particle chain with a chain containing as many spins as numerically possible only. In fig.~\ref{fig:centralsipn-chain-meanfield} and fig.~\ref{fig:centralsipn-chain-mpc}, the dynamics of a $20001$ particle mean-field simulation and a $401$ particle MPC simulation are used as the best possible approximation of an infinite chain for three different spin-spin distances, respectively.
In most cases the addition of further spins affects the central spin less and less and is approximately linear in this double logarithmic plot, i.e.~the trace distance between the infinite chain and the N-particle chain for a certain distance can be estimated by $T(N,\inf)=C_d N^k_d$. By fitting this function to the numerical results we can determine the exponent $k_d$ and the factor $C_d$ depending on the distance which is plotted in fig.~\ref{fig:centralsipn-chain-tdestimate}.
For nearly all distances both the mean-field method as well as the MPC method predict that adding further particles has an effect proportional to $\frac{1}{N}$ only, allowing us to easily estimate the number of particles needed to approximate the infinite chain dynamics to a desired accuracy. However, when the spin-spin distance is close to a multiple of the transition wavelength $\lambda_0$, the dynamics of the infinite chain never seems to be captured by a finite size approximation.

\subsection{Hexagonal lattice}
\begin{figure*}[ht]
  \includegraphics{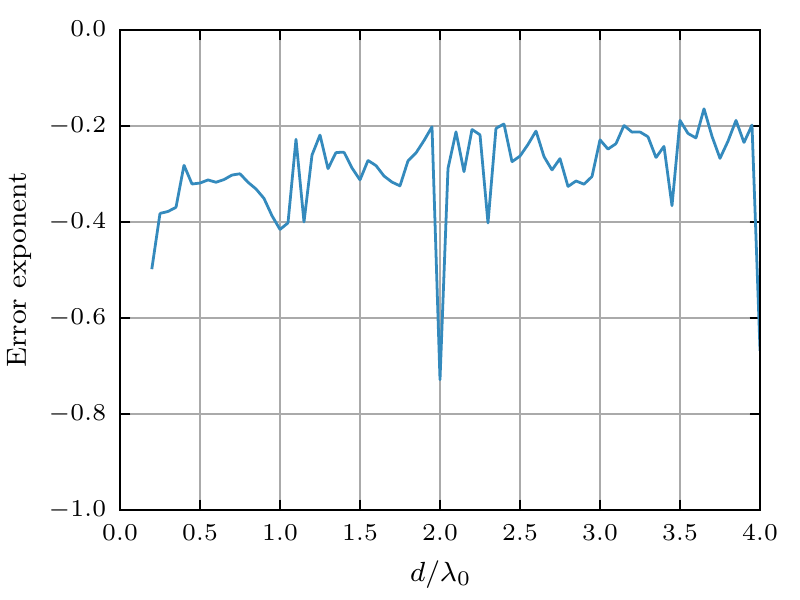}
  \includegraphics{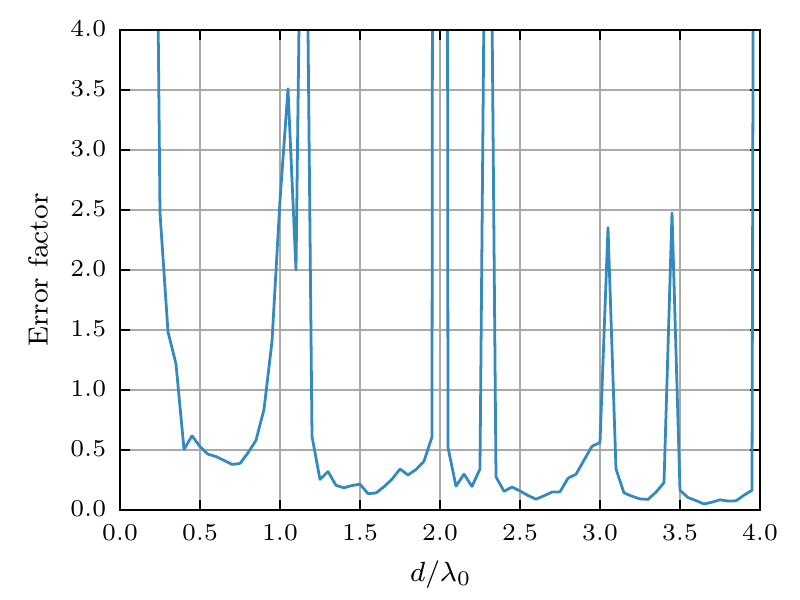}
  \caption{Error exponent $k_d$ and error factor $C_d$ dependence on the spin-spin distance determined in the previous fits $C_d*N^{k_d}$ using mean-field in the case of a hexagonal lattice.}
  \label{fig:centralspin-hexagonallattice-tdestimate}
\end{figure*}

With these very encouraging results for a 1D chain, let us see, if this holds true for higher dimensional geometries as well. Unfortunately we failed to obtain convincing results for 3D cubic lattices, since the number of particles needed for convergence of the numerical result turns out to very much exceed the possibility of MPC and even of the mean-field method. For 2D geometries at least, the mean-field method delivers some meaningful, albeit by far not as beautiful results. Fig.~\ref{fig:centralspin-hexagonallattice-tdestimate} shows the numerically obtained approximations for the error exponent and the error factor for the case of a hexagonal lattice, where additional particles are added in rings around the central spin.
The outcome looks rather noisy, probably due to too small a choice for the number of particles used as an approximation of the infinite lattice. It turns out that compared to the chain a lot more particles are needed to reliably approximate an infinite hexagonal lattice, i.e.~the influence of additional particles reduces the error with approximately $N^{-0.3}$, where this exponent is a rather rough estimate.

\section{Numerical complexity of the different methods}
\begin{figure}[ht]
  \includegraphics{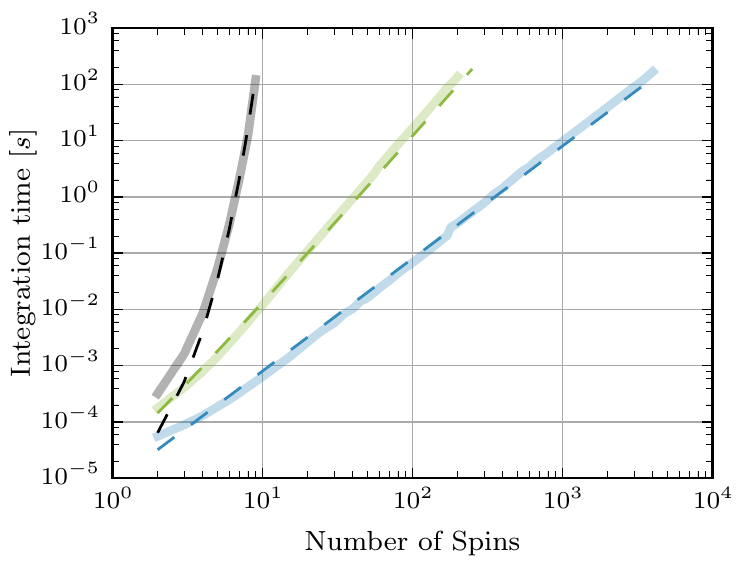}
  \caption{Time needed to integrate a spin chain consisting of $N$ spins from $0$ to $\gamma^{-1}$ on a single CPU. The solid lines are the result of the benchmarks for master (black), MPC (green) and mean-field (blue), the dashed lines are the corresponding theoretical predictions.}
  \label{fig:benchmark}
\end{figure}

Finally we want to add some considerations on the memory and CPU requirements of the different methods and show that our implementations behave as expected in this regard. When solving the master equation the state of the system is captured as a density matrix of dimension $2^{2N}$. The time evolution according to a master equation is equivalent to a matrix-matrix multiplication and therefore has a time complexity of $O(2^{3N})$. In the case of mean-field a state can be characterized by $3N$ real numbers and according to the mean-field equations (eq.~\ref{eq:mf-timeevolution}) the time complexity is then approximately $O(N^2)$. For the MPC method the state consists of one mean-field state and nine correlation matrices of the form $C^{\alpha\beta}_{ij} = \expect{\sigma_i^\alpha \sigma_j^\beta}$. Using the relation $C^{\alpha\beta}_{ij}=C^{\alpha\beta}_{ji}$ means that we need roughly $\frac{9N^2}{2}$ real numbers to represent one MPC state. The time complexity is, according to the MPC equations (eq.~\ref{eq:mpc-timeevolution}), approximately $O(N^3)$. The results of this analysis are presented in fig.~\ref{fig:benchmark}.

\section{Conclusions and outlook}
We have demonstrated that an effective mean-field method with added pair correlations provides a numerically efficient and surprisingly accurate method to simulate open spin systems with general non local spin-spin interaction and collective decay up to moderately high particle numbers and significant interaction strength. Particularizing to dipole-dipole interaction and collective spontaneous decay has allowed us to establish a numerical estimate of the accuracy and scaling properties of our methods. Furthermore we can show for 1D chains that tractable system sizes already approach the behavior of infinite systems allowing for an estimate of the magnitude of the error due to the truncation of the system. For 2D systems the lowest order mean-field approach still allows to reach adequate system sizes to approximate infinite systems, whereas the scaling is unfavorable to accurately approximate infinite 3D systems. In future work, we plan to apply these methods to study collectively enhanced as well as suppressed decay in magic wavelength lattices for clock atoms. The simulations should also provide us with predictions of geometries and excitation schemes to minimize dipole-dipole induced shifts in order to improve the accuracy of atomic clocks. Possible approaches would be to analyze different geometries, use initial phase spread rotations and spin squeezing. As an interesting extension of this model we also want to embed such spin systems inside a cavity and derive corresponding mean-field and MPC equations for the arising infinite range interactions. This should give us a basis to simulate super-radiant lasers for larger ensembles including their interaction. Note, that as we are simulating an open system anyway, including a finite bath temperature will hardly change the complexity of these calculations and could be used to identify temperature dependent phase transitions in the system.

\begin{acknowledgement}
This work has been supported by the Austrian Science Fund FWF through the SFB F4013 FoQuS.
\end{acknowledgement}

\bibliographystyle{epj}
\bibliography{references-meanfieldmethod}

\end{document}